\documentclass[10pt,journal,twocolumn]{IEEEtran}
\IEEEoverridecommandlockouts
\usepackage{cite}
\usepackage{amsmath,amssymb,amsfonts,booktabs}
\usepackage{algorithmic}
\usepackage{graphicx}
\usepackage{textcomp}
\usepackage{xcolor}
\usepackage{enumerate}
\usepackage{epstopdf}
\usepackage{array}
\usepackage{easyReview}
\usepackage{textcomp}
\usepackage{stfloats}
\usepackage{url}
\usepackage{bm}
\usepackage{amsthm}
\usepackage{amsthm}

\usepackage{algorithm}
\usepackage{algorithmic}
\allowdisplaybreaks[4]

\usepackage{array}
\usepackage{stfloats}
\usepackage{cuted}
\usepackage{url}
\usepackage{verbatim}

\newtheorem{theorem}{Theorem}
\newtheorem{remark}{Remark}
\newtheorem{proposition}{Proposition}

\usepackage{color}

\begin{document}

\title{RSMA-Enhanced Data Collection in RIS-Assisted Intelligent Consumer Transportation Systems}

\author{
        Chunjie Wang,
        Xuhui Zhang,
        Wenchao Liu,
        Jinke Ren,
        Shuqiang Wang,~\IEEEmembership{Senior Member,~IEEE,}
        Yanyan Shen,
        Kejiang Ye,~\IEEEmembership{Senior Member,~IEEE,}
        and Kim Fung Tsang,~\IEEEmembership{Fellow,~IEEE}

\thanks{Chunjie Wang is with Shenzhen Institutes of Advanced Technology, Chinese Academy of Sciences, Shenzhen 518055, China, and also with University of Chinese Academy of Sciences, Beijing 100049, China (e-mail: cj.wang@siat.ac.cn)
}

\thanks{
Xuhui Zhang and Jinke Ren are with the Shenzhen Future Network of Intelligence Institute, the School of Science and Engineering, and the Guangdong Provincial Key Laboratory of Future Networks of Intelligence, The Chinese University of Hong Kong, Shenzhen, Guangdong 518172, China (e-mail: xu.hui.zhang@foxmail.com; jinkeren@cuhk.edu.cn).
}

\thanks{
Wenchao Liu is with the School of Automation and Intelligent Manufacturing, Southern University of Science and Technology, Shenzhen 518055, China (e-mail: wc.liu@foxmail.com).
}

\thanks{
Shuqiang Wang, Yanyan Shen, Kejiang Ye, and Kim Fung Tsang are with Shenzhen Institutes of Advanced Technology, Chinese Academy of Sciences, Guangdong 518055, China (e-mail: sq.wang@siat.ac.cn; yy.shen@siat.ac.cn; kj.ye@siat.ac.cn; kftsang@ieee.org)
}

}

\maketitle

\begin{abstract}
This paper investigates the data collection enhancement problem in a reconfigurable intelligent surface (RIS)-empowered intelligent consumer transportation system (ICTS). We propose a novel framework where a data center (DC) provides energy to pre-configured roadside unit (RSU) pairs during the downlink stage. While in the uplink stage, these RSU pairs utilize a hybrid rate-splitting multiple access (RSMA) and time-division multiple access (TDMA) protocol to transmit the processed data to the DC, while simultaneously performing local data processing using the harvested energy. Our objective is to maximize the minimal processed data volume of the RSU pairs by jointly optimizing the RIS downlink and uplink phase shifts, the transmit power of the DC and RSUs, the RSU computation resource allocation, and the time slot allocation. To address the formulated non-convex problem, we develop an efficient iterative algorithm integrating alternating optimization and sequential rank-one constraint relaxation methods. Extensive simulations demonstrate that the proposed algorithm significantly outperforms baseline schemes under diverse scenarios, validating its effectiveness in enhancing the data processing performance for intelligent transportation applications.
\end{abstract}


\section{Introduction}

\IEEEPARstart {W}{ith} the rapid advancement of information and communication technologies, the Internet of Things (IoT) has emerged as a transformative paradigm, connecting physical devices and enabling intelligent decision-making in diverse fields such as smart cities, healthcare systems, and industrial automation \cite{10430225,10552149}. 
As IoT technologies continue to evolve, their applications have become increasingly specialized, leading to significant advancements in Internet of Vehicles (IoV) for intelligent consumer transportation systems (ICTSs). 
Integrating with advanced communications, processing, and sensor technologies, traffic efficiency, safety, and sustainability are thus enhanced \cite{10384717,10904066}. 
Meanwhile, data collection plays a fundamental role in ensuring the operational efficiency and analytical capabilities for the ICTS. By leveraging real-time data from vehicles and infrastructures, ICTS not only improves the driving experience but also contributes to smarter urban planning \cite{10393515}.
Effective data collection underpins the performance and scalability of modern ICTS, facilitating informed decision-making and responsive management strategies.

However, due to the complexity of urban traffic environments, there are many obstacles between communication links, which can lead to a decrease in the efficiency of data collection in the ICTS. 
Therefore, how to efficiently collect data in the ICTS remains a challenging task.
To address this challenge, recent advances in wireless communication technologies have opened new possibilities for improving data collection efficiency in the ICTS. Among these, the emergence of reconfigurable intelligent surface (RIS), also known as intelligent reflecting surface, has shown great potential in enhancing signal coverage, reducing transmission latency, and improving overall network reliability \cite{9956000, wang2025joint}. 
By intelligently reconfiguring the propagation environment, RIS can help overcome communication barriers caused by dynamic traffic conditions and infrastructure limitations \cite{10533201}. 
This makes RIS a promising enabler for more efficient and robust data collection in the ICTS, especially in complex urban transportation scenarios.

While RIS offers promising improvements in wireless connectivity for the ICTS, several limitations remain in their application to data collection tasks. 
One key challenge is the difficulty in efficiently managing multiple concurrent data streams from distributed sensors and vehicles, especially under high mobility and heterogeneous channel conditions. 
Traditional multiple access schemes often struggle to provide reliable and scalable performance in such environments, limiting the full potential of the RIS-enhanced communication \cite{10133841}. 
To overcome these constraints, recent research has highlighted the benefits of rate-splitting multiple access (RSMA), a novel non-orthogonal transmission strategy that enables flexible interference management and improved spectral efficiency \cite{9831440}.
Specifically, the RSMA splits user data into common and private components, which are transmitted using linear precoding, and the receiver decodes them sequentially via successive interference cancellation \cite{9832611}. 

Integrating RSMA with the RIS-assisted architectures presents a compelling opportunity to enhance data collection capabilities in the ICTS, particularly in scenarios involving large-scale and high-density vehicular communications. Specifically, the RIS can intelligently reshape the propagation environment to improve signal reliability, while RSMA manages multi-user interference through its innovative non-orthogonal transmission paradigm. Their combination not only enhances link robustness but also improves spectral efficiency. However, the core challenge of this integration lies in the complex interaction between the RIS phase shift optimization and the RSMA-based resource allocation. Under constraints such as energy harvesting and data transmission, this interaction gives rise to highly coupled and complex non-convex optimization problems. As a result, there is a lack of effective methods to maximize the system performance.

Therefore, motivated by above, we propose a novel RIS-empowered data collection framework for the ICTS.
In the considered ICTS, a data center (DC) conducts downlink wireless energy transfer (WET) to multiple roadside unit (RSU) pairs deployed on both sides of the road. These RSUs then utilize the harvested energy to transmit the processed data to the DC and performing local data processing. A RIS is employed to enhance both downlink WET and uplink data transmission processes. To mitigate interference and enhance transmission efficiency, a hybrid protocol merging RSMA with time-division multiple access (TDMA) is adopted. Specifically, TDMA is used across different RSU pairs to avoid inter-pair interference, while RSMA is applied within each pair to improve transmission efficiency. Focusing on data collection, this work formulates an optimization problem aiming to maximize the minimal processed data volume across all RSU pairs, subject to the system time limitations, transmit power constraints, RIS phase shift constraints, and constraints on computation resources in CPU cycles (CRCs). To solve this problem, an efficient algorithm integrating alternating optimization (AO) and sequential rank-one constraint relaxation (SROCR) methods is proposed.

The main contributions of this work can be summarized as follows:
\begin{itemize}
    \item We propose a comprehensive optimization problem that jointly designs time slot allocation, transmit power control, RIS phase shift coefficients, and CRCs to maximize the minimal processed data in the RIS-assisted ICTS. In addition, a hybrid RSMA and TDMA protocol is proposed to tackle the interference issues and enhance data transmission efficiency.
    \item The formulated problem is inherently non-convex and computationally challenging. To tackle this, we first transform the original problem and decompose it into four subproblems. By leveraging the AO approach, we develop an iterative algorithm that efficiently solves each subproblem while ensuring convergence, providing a feasible solution for the original problem.
    \item  We performed comprehensive simulation experiments to assess the performance of the proposed scheme, and the results show that it significantly outperforms baseline schemes in improving the data processing capabilities of the ICTS.
\end{itemize}

\textit{Organizations:} The rest of this paper is organized as follows.
Section II presents the related works.
Section III introduces the ICTS and formulates the minimal processed data maximization problem.
In Section IV, we develop an AO algorithm to address the problem, along with an assessment of its convergence and complexity.
Section V illustrates the simulation results demonstrating the performance gains of the proposed scheme over various baseline schemes.
The paper is concluded in Section VI.

\textit{Notations:} The notations mentioned in this paper are introduced below.
$ {{\mathbb{C}}^{M\times N}} $ denotes the $ M \times N $ complex matrix $ \mathbb{C} $. $ {\cal C}{\cal N}(\mu ,{\sigma ^2}) $ denotes the circularly symmetric complex Gaussian distribution with $ \mu $ mean and $ {\sigma ^2} $ variance.
$ \mathrm{j} $ represents the imaginary unit, where $ {\mathrm{j}^2} = -1 $. For a generic matrix $ {\boldsymbol{G}} $, $ {{\boldsymbol{G}}^{\mathsf{H}}} $, $ {{\boldsymbol{G}}^{\mathsf{T}}} $, $ \mathrm{tr}({\boldsymbol{G}}) $, and $\mathsf{rank}({\boldsymbol{G}})$ denote the conjugate transpose, transpose, trace, and rank of $ {\boldsymbol{G}} $, respectively. $ |\cdot| $ represents the absolute value.

\section{Related Works}

In this section, we summarize the related works on three critical areas: data collection systems, the RIS-empowered systems, and the RSMA-assisted communications. These works collectively establish the theoretical and practical basis for our proposed system.

\subsection{Data Collection}

ICTSs rely on real-time traffic data provided by IoV to conduct analysis and support decision-making processes, with the objective of optimizing traffic flow and reducing traffic accidents. Therefore, data collection plays a crucial role in the effectiveness and performance of the ICTSs.
The authors in \cite{10437121} proposed an autonomous aerial vehicle (AAV)-assisted vehicular network data collection framework based on real-time traffic scenarios, which maximizes the success rate of data transmission by jointly optimizing AAV flight trajectories, access decisions, and resource allocation.
The authors in \cite{10365757} proposed an AAV-assisted system architecture for vehicular data uploading, and developed an algorithm based on deep reinforcement learning (DRL) aimed at maximizing the amount of collected data through trajectory design.
The authors in \cite{9989729} employed a Tabu search-based algorithm to address the AAV path planning problem for large-scale and continuous data collection from heterogeneous IoT devices.
The authors in \cite{10233692} minimized the completion time while ensuring that ground nodes perform uplink data transmission within their energy budgets, decomposing the problem into three subproblems for efficient solution.
The authors in \cite{9875063} jointly optimized the AAV's three-dimensional trajectory and the transmission scheduling of sensor nodes to achieve secure and energy-efficient data collection.
However, existing data collection approaches related to the ICTS are relatively simplistic and suffer from low efficiency to some extent, as they have not integrated advanced technologies such as RIS to enhance the efficiency of data collection.

\subsection{RIS}

With the growing interest in intelligent and controllable wireless environments, a significant body of research has focused on RIS, which enhances the performance of wireless communication systems by smartly reconfiguring the propagation environment.
The authors in \cite{10261304} considered a RIS-assisted IoV network and employed a Markov game-based multi-agent DRL algorithm, where multiple agents interact with the environment to collaboratively optimize decisions toward maximizing system utility.
The authors in \cite{10663259} employed a DRL-based soft actor-critic algorithm to optimize RIS-assisted vehicular network communications.
The authors in \cite{10381620} proposed a RIS-assisted vehicle-to-vehicle communication system and developed an efficient algorithm, which demonstrates superior effectiveness and application potential in RIS-assisted vehicular networks under various network conditions compared to baseline schemes.
The authors in \cite{10129124} focused on the problem of deploying a RIS-enabled AAV to provide aerial line-of-sight communication services for IoV.
The authors in \cite{10520918} proposed a system model integrating RIS with non-orthogonal multiple access (NOMA), where a large language model was utilized to optimize resource block allocation and decoding order, followed by sequential optimization of initial phase shift, power control, and phase shift adjustment.
Although some existing studies have considered RIS-assisted IoV, they do not address how to mitigate the interference among communications, leaving ample room for further exploration in this area.

\subsection{RSMA}

Under the background of advanced multiple access schemes for the next-generation wireless networks, RSMA has become a promising technology that can effectively handle communication interference.
The authors in \cite{10681863} focused on space-air-ground-sea networks and proposed a downlink multi-user coherent beamforming RSMA system to enhance system performance.
The authors in \cite{10269133} established an RSMA-AAV-assisted downlink communication system model, enabling the AAV to serve two ground users simultaneously, and formulated a joint non-convex optimization problem involving communication scheduling, common rate allocation, transmit power control, and AAV trajectory design to maximize the AAV's energy efficiency.
The authors in \cite{9952717} aimed to address the problem of base station transmit power allocation in a downlink RSMA system under the stochastic characteristics of wireless channels, user mobility, and energy harvesting, with the objective of maximizing the long-term total system rate.
The authors in \cite{10330667} considered a single-input single-output uplink RSMA transmission system and formulated an optimization problem to minimize the maximum transmission delay, where a strong-weak user pairing strategy was adopted to reduce complexity by transforming the original problem into a bandwidth and power allocation optimization.
The authors in \cite{10200653} constructed a system model that incorporates the uplink training phase for channel estimation and RSMA-assisted uplink data transmission, and formulated a power allocation problem based on maximizing the minimal spectral efficiency.
The authors in \cite{10393781} focused on the joint power and precoder design in energy-harvesting aerial networks to maximize the long-term total data rate, where simulation results highlight the significance of power allocation strategies and RSMA in enhancing network performance.
While RSMA has been explored in various domains, research efforts dedicated to its applications in the ICTS remain limited, indicating significant potential for leveraging RSMA in the ICTS.

\section{System Model and Problem Formulation}

\begin{figure}[!htbp]
	\centering
	\includegraphics[width=0.82\linewidth]{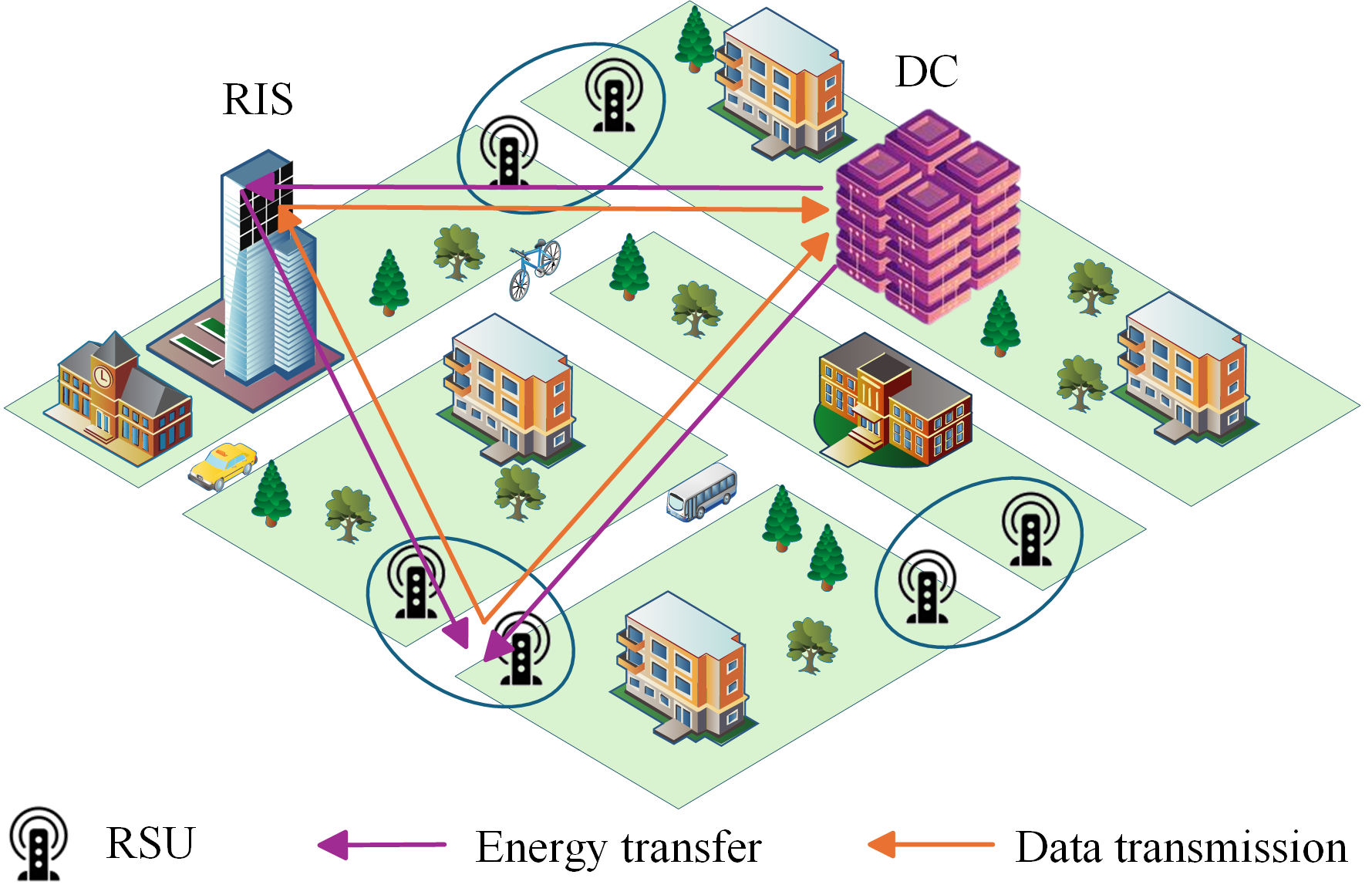}
	\caption{The RIS-empowered data collection for the ICTS with RSMA.}
	\label{fig:ITS_model}
\end{figure}

As illustrated in Fig. \ref{fig:ITS_model}, we consider an ICTS consists of one single-antenna DC, $2 M$ single-antenna RSUs which are assigned as $M$ pairs on both sides of the road, and a RIS with $N$ reflecting elements.
The set of the RSU pairs is given by $\mathcal{M} = \{1,2,\ldots,M\}$, and the set of the RIS reflecting elements is given by $\mathcal{N} = \{1,2,\ldots,N\}$.
Let $m_i$ denote the $i$-th RSU in the $m$-th pair, where $i \in \{1,2\}$.

We consider a novel hybrid RSMA and TDMA protocol.
Within each RSU pair, the RSMA is utilized to facilitate the uplink data transmission of the RSUs,
while the TDMA is employed across different pairs to eliminate interference.
The duration of time interval is $T$ seconds, which is divided into $M+1$ time slots, indexed by $\mathcal{T} = \{0,1,2,\ldots, M\}$, where the first time slot $t_0$ represents the downlink energy transfer time by the DC, and the rest of $M$ time slots are allocated to the $M$ pairs of the RSUs by the order of $\mathcal{M}$, respectively.

\subsection{Wireless Energy Transfer Model}
Let $g_{m,i} \in \mathbb{C}^{1\times 1}$, $\boldsymbol{G}  \in \mathbb{C}^{N\times 1}$, and $\boldsymbol{h}_{m,i}  \in \mathbb{C}^{N\times 1}$ denote the channel gains from the DC to the RSU $m_i$, from the DC to the RIS, and from the RIS to the RSU $m_i$, respectively. Meanwhile, the phase shift coefficient vector of the RIS during the WET is $\boldsymbol{\theta}_0 = [\theta_{0,1}, \theta_{0,2},\ldots, \theta_{0,N} ]^{\mathsf{T}}$, where $\theta_{0,n} \in [0,2\pi)$, $\forall n \in \mathcal{N}$.
We assume that the reflection coefficients are normalized to $1$ \cite{9133107, 9417469}.
Hence, the phase shift coefficient matrix of RIS during WET can be given by $\boldsymbol{\Theta}_0 = \mathsf{diag} \{ \mathrm{e}^{\mathrm{j}\theta_{0,1}},\mathrm{e}^{\mathrm{j}\theta_{0,2}},\ldots,\mathrm{e}^{\mathrm{j}\theta_{0,N}} \}$, where $\mathrm{j}$ represents the imaginary unit.
The received radio frequency power at the RSU $m_i$ can be expressed as
\begin{equation}
    p_{m,i}^{\mathrm{down}} = p_0 \vert \boldsymbol{h}_{m,i}^{\mathsf{H}} \boldsymbol{\Theta}_0 \boldsymbol{G} + g_{m,i} \vert^2,
    \label{recepow}
\end{equation}
where $p_0$ represents the transmit power of the DC.
We consider a practical non-linear energy harvesting (NLEH) model \cite{9930110, 10288203}.
The harvested energy of the RSU $m_i$ can be given by
\begin{equation}
    e_{m,i}^{\mathrm{NLEH}} = t_0
    \frac{
    \Phi_{m,i} - \Lambda_{m,i} \Omega_{m,i}
    }{
    1-\Omega_{m,i}
    },
\end{equation}
where $\Lambda_{m,i}$ denotes the maximum harvestable power at the RSU $m_i$. $\Omega_{m,i}$ characterizes the zero-input/zero-output response of the NLEH, and is given by\cite{7264986}
\begin{equation}
\Omega_{m,i} = \frac{1}{
    1 + \mathrm{e}^{a_{m,i}b_{m,i}}},
\end{equation}
where $a_{m,i}$ and $b_{m,i}$ are constants determined by the specific circuit parameters of RSU $m_i$. Meanwhile, $\Phi_{m,i}$ denotes the logistic function with respect to the received power $p_{m,i}^{\mathrm{down}}$, which is given by
\begin{equation}
    \Phi_{m,i} = \frac{
    \Lambda_{m,i}
    }{1+
    \mathrm{e}^{-a_{m,i}(p_{m,i}^{\mathrm{down}}-b_{m,i})}
    }.
\end{equation}

\subsection{Data Processing and Transmission Models}
The RSUs should process their sensed data first. Let $F_{m,i}$ denote the scheduled computation resources in CRCs of the RSU $m_i$, and $f_{m,i}$ represent the CRCs required to process one bit sensed data according to the circuit physical characteristic of the RSU $m_i$.
Hence, the data processing rate of the RSU $m_i$ can be given by 
\begin{equation}
r_{m,i} = \frac{F_{m,i}}{f_{m,i}}.
\end{equation}
We assume that the RSUs utilize the rest of the time interval for data processing after harvesting energy from the DC. Therefore, the processable data volume is $d_{m,i} = r_{m,i} (T-t_0)$.
Meanwhile, the energy consumption for data processing of the RSU $m_i$ can be expressed as
\begin{equation}
    e_{m,i}^{\mathrm{proc}} = \kappa F_{m,i}^3 (T-t_0),
\end{equation}
where $\kappa$ denotes the circuit capacitance coefficient \cite{7524497, cyberzhang23}.

In addition to data processing, the RSUs also transmit the processed data to the DC. We assume that each RSU is equipped with a caching module, where the processed data from the previous time interval is stored and held for transmission.
The RSUs within each pair transmit data with RSMA protocol. Accordingly, the transmit signal of the RSU $m_i$ is divided into two parts, i.e., $x_{m,i,1}$ and $x_{m,i,2}$, and the allocated transmit power for each part are given by $\rho p_{m,i}$ and $(1-\rho) p_{m,i}$, respectively, where $p_{m,i}$ is the total uplink transmit power of the RSU $m_i$.
Besides, the transmit signal of the RSU $m_j, j \neq i$, which is within the same pair of the RSU $m_i$, should be directly transmitted to the DC.

Let $\Tilde{g}_{m,i} \in \mathbb{C}^{1\times 1}$, $\Tilde{\boldsymbol{G}}  \in \mathbb{C}^{N\times 1}$, and $\Tilde{\boldsymbol{h}}_{m,i}  \in \mathbb{C}^{N\times 1}$ denote the uplink channel gains from the RSU $m_i$ to the DC, from the RIS to the DC, and from the RSU $m_i$ to the RIS, respectively. The phase shift coefficient vector of the RIS for assisting the uplink transmission of the $m$-th pair RSUs is given by $\boldsymbol{\theta}_m = [\theta_{m,1}, \theta_{m,2},\ldots, \theta_{m,N} ]^{\mathsf{T}}$, where $\theta_{m,n} \in [0,2\pi)$, $\forall n \in \mathcal{N}$.
Similarly, the reflection coefficients during the uplink transmission are also normalized to $1$. Then, the phase shift coefficient matrix of RIS for assisting the $m$-th pair RSUs is given by $\boldsymbol{\Theta}_m = \mathsf{diag} \{ \mathrm{e}^{\mathrm{j}\theta_{m,1}},\mathrm{e}^{\mathrm{j}\theta_{m,2}},\ldots,\mathrm{e}^{\mathrm{j}\theta_{m,N}} \}$.
As a result, the received signal transmitted by the $m$-th RSU pair at the DC can be expressed as
\begin{equation}
\begin{split}
    y_m =&\ \sqrt{\rho p_{m,i}} \left (
        \Tilde{g}_{m,i} + \Tilde{\boldsymbol{G}}^{\mathsf{H}} \boldsymbol{\Theta}_m \Tilde{\boldsymbol{h}}_{m,i}
    \right ) x_{m,i,1} \\
    &+ \sqrt{(1-\rho) p_{m,i}} \left (
        \Tilde{g}_{m,i} + \Tilde{\boldsymbol{G}}^{\mathsf{H}} \boldsymbol{\Theta}_m \Tilde{\boldsymbol{h}}_{m,i}
    \right ) x_{m,i,2} \\
    &+ \sqrt{ p_{m,j}} \left (
        \Tilde{g}_{m,j} + \Tilde{\boldsymbol{G}}^{\mathsf{H}} \boldsymbol{\Theta}_m \Tilde{\boldsymbol{h}}_{m,j}
    \right ) x_{m,j} + n_m,\\
    &\quad\quad\quad\quad\quad\quad\quad\quad\quad\quad\quad i,j \in \{1,2\},\ i \neq j,
\end{split}
\end{equation}
where $n_m \sim \mathcal{C}\mathcal{N} (0, \sigma_m^2)$ denotes the noise at the DC. By the decoding order $x_{m,i,1} \rightarrow x_{m,j} \rightarrow x_{m,i,2}$ \cite{11086503,10032159}, the SINR for $x_{m,i,1}$ is given at the top of next page, and the SINR for $x_{m,j}$ and $x_{m,i,2}$ can be given by
\begin{figure*}[!t]
\begin{equation}
    \gamma_{m,i,1} = \frac{
    \rho p_{m,i} \left \vert
        \Tilde{g}_{m,i} + \Tilde{\boldsymbol{G}}^{\mathsf{H}} \boldsymbol{\Theta}_m \Tilde{\boldsymbol{h}}_{m,i}
    \right \vert^2
    }{(1-\rho) p_{m,i} \left \vert
        \Tilde{g}_{m,i} + \Tilde{\boldsymbol{G}}^{\mathsf{H}} \boldsymbol{\Theta}_m \Tilde{\boldsymbol{h}}_{m,i}
    \right \vert^2 + 
    p_{m,j} \left \vert
        \Tilde{g}_{m,j} + \Tilde{\boldsymbol{G}}^{\mathsf{H}} \boldsymbol{\Theta}_m \Tilde{\boldsymbol{h}}_{m,j}
    \right \vert^2 + \sigma_m^2},
\end{equation}
\hrulefill
\end{figure*}
\begin{equation}
    \gamma_{m,j} = \frac{
     p_{m,j} \left \vert
        \Tilde{g}_{m,j} + \Tilde{\boldsymbol{G}}^{\mathsf{H}} \boldsymbol{\Theta}_m \Tilde{\boldsymbol{h}}_{m,j}
    \right \vert^2
    }{(1-\rho) p_{m,i} \left \vert
        \Tilde{g}_{m,i} + \Tilde{\boldsymbol{G}}^{\mathsf{H}} \boldsymbol{\Theta}_m \Tilde{\boldsymbol{h}}_{m,i}
    \right \vert^2 + \sigma_m^2},
\end{equation}
\begin{equation}
    \gamma_{m,i,2} = \frac{
     (1-\rho)p_{m,i} \left \vert
        \Tilde{g}_{m,i} + \Tilde{\boldsymbol{G}}^{\mathsf{H}} \boldsymbol{\Theta}_m \Tilde{\boldsymbol{h}}_{m,i}
    \right \vert^2
    }{\sigma_m^2}.
\end{equation}
Accordingly, the data transmission rate of the RSU $m_i$ and the RSU $m_j$ can be given by
\begin{equation}
    r_{m,i}^{\mathrm{tr}} = B\left( \log_2(1+\gamma_{m,i,1}) + \log_2(1+\gamma_{m,i,2}) \right),
\end{equation}
\begin{equation}
    r_{m,j}^{\mathrm{tr}} = B \log_2(1+\gamma_{m,j}),
\end{equation}
where $B$ is the system bandwidth. Then, the transmitted data volume and the corresponding energy consumption of the RSU $m_i$ and the RSU $m_j$ are given by
\begin{equation}
    d_{m,i}^{\mathrm{tr}} = t_m r_{m,i}^{\mathrm{tr}},\ 
    d_{m,j}^{\mathrm{tr}} = t_m r_{m,j}^{\mathrm{tr}},
\end{equation}
\begin{equation}
    e_{m,i}^{\mathrm{tr}} = t_m p_{m,i},\ 
    e_{m,j}^{\mathrm{tr}} = t_m p_{m,j}.
\end{equation}

\subsection{Problem Formulation}
We aim to maximize the minimal processed data of each pair of RSUs within the time interval, by optimizing the time slot allocation, $\boldsymbol{t} = \{ t_m, m \in \mathcal{T}\}$, the transmit power of the DC, $p_0$, the transmit power of each RSU, $\boldsymbol{p}= \{ p_{m,i}, m \in \mathcal{M}, i \in \{1,2\} \}$, the RIS phase shift coefficients, $\Bar{\boldsymbol{\Theta}} = \{ \boldsymbol{\Theta}_0\ \text{and}\ \boldsymbol{\Theta}_m, m \in \mathcal{M} \}$, and the CRCs of each RSU, $\boldsymbol{F} = \{ F_{m,i}, m \in \mathcal{M}, i \in \{1,2\}\}$. Let $\boldsymbol{X} = \{ \boldsymbol{t} , p_0, \boldsymbol{p}, \Bar{\boldsymbol{\Theta}},\ \text{and}\ \boldsymbol{F}\}$ denote the variable set. Then, the problem is formulated as
\begin{subequations} { 
\begin{flalign}
 (\textbf{P1}):\ \max_{\boldsymbol{X}} \quad & \min_{m}\ D_m  \nonumber\\
 {\rm{s.t.}}  \quad 
  &\sum_{m=0}^M t_m \leq T,\label{p1a}\\
 &0 \leq t_m \leq T,\ \forall m \in \mathcal{T},\label{p1b}\\
 & 0 \leq p_0 \leq P_{\max},\label{p1c}\\
 & 0 \leq p_{m,i} \leq P_{m, i, \max},\ \forall m\in \mathcal{M},\ \forall i \in \{1,2 \},  \label{p1d}\\
 &e_{m,i}^{\mathrm{proc}} + e_{m,i}^{\mathrm{tr}} \leq e_{m,i}^{\mathrm{NLEH}}, \nonumber\\
 &\quad\quad\forall m \in \mathcal{M},\ \forall i \in \{1,2 \},\label{p1e}\\
 & 0\leq \theta_{0,n} \le 2\pi,\ \forall n \in \mathcal{N},\label{p1f}\\
 &0\leq \theta_{m,n} \le 2\pi,\ \forall n \in \mathcal{N},\ \forall m \in \mathcal{M},\label{p1g}\\
& 0 \leq F_{m,i} \leq F_{m, i, \max},\ \forall m\in \mathcal{M},\ \forall i \in \{1,2 \},  \label{p1h}
\end{flalign}}\end{subequations}where $D_m$ denotes the total processed data volume of the $m$-th RSU pair during the interval, which is given by
\begin{equation}
    D_m = \sum_{i=1}^2 \left(d_{m,i}^{\mathrm{}} + d_{m,i}^{\mathrm{tr}}\right), \label{dm}
\end{equation}
and $P_{\max}$ represents the maximum energy transmit power of the DC, $P_{m,i,\max}$ is the maximum transmit power of the RSU $m_i$, and $F_{m,i,\max}$ denotes the maximum CRCs of the RSU $m_i$.
Constraints \eqref{p1a} and \eqref{p1b} represent the time limitations for energy transfer and data processing.
Constraint \eqref{p1c} and \eqref{p1d} denote that the transmit power of the DC and each RSU cannot exceed their power bounds.
Constraint \eqref{p1e} represents that the energy consumption of each RSU cannot exceed its total harvested energy.
Constraints \eqref{p1f} and \eqref{p1g} are the RIS phase constraints.
Constraint \eqref{p1h} denotes that the allocated CRCs of each RSU cannot exceed its maximum CRCs.

\section{Joint Optimization of the Time Allocation, the Transmit Power, the RIS, and the CRCs}
\subsection{Problem Analysis}
Firstly, problem \textbf{P1} can be proved as a non-convex optimization problem through Theorem \ref{p1nonvex} as follows.
\begin{theorem} \label{p1nonvex}
\textbf{P1} is a non-convex optimization problem.
\begin{proof}
Firstly, the objective function is formulated in a \textsc{max-min} form, which is inherently non-convex. Moreover, the energy constraint in \eqref{p1e} is also non-convex. At the same time, the signal transmission rate depends on the SINR—a complex fractional expression involving multiple coupled variables. Furthermore, the phase shift variables of the RIS appear in the channel gain in the form of complex exponentials, introducing additional non-convexity. As a result, \textbf{P1} constitutes a non-convex optimization problem.

\end{proof}
\end{theorem}
Accordingly, problem \textbf{P1} is hard to be solved by conventional methods.
To solve it, we propose an AO-based algorithm to optimize a part of the optimization variables while fixed others in an alternating manner.

\subsection{Problem Transformation}
To transform the \textsc{max-min} objective function of problem \textbf{P1}, we introduce an auxiliary variable $\ell$ to reformulate problem \textbf{P1} as \textbf{P2}, which is given by
\begin{subequations} { 
\begin{flalign}
 (\textbf{P2}):\ \max_{\boldsymbol{X}, \ell} \quad & \ell  \nonumber\\
 {\rm{s.t.}}  \quad 
  &\eqref{p1a}-\eqref{p1h},\\
 &\ell \leq D_m,\ \forall m \in \mathcal{M}.\label{p2b}
\end{flalign}}\end{subequations}
Meanwhile, the processed data volume of each RSU pair can be reformulated through the following proposition.
\begin{proposition}
The processed data volume of the $m$-th RSU pair can be rewritten as
\begin{equation}
    D_m = \sum_{i=1}^2 d_{m,i} + 
    Bt_m \left(
        \log_2 \left(
            1+\frac{\Bar{p}_{m,i} + \Bar{p}_{m,j}}{\sigma_m^2}
        \right)
    \right), \label{dm2}
\end{equation}
where
\begin{equation}
    \Bar{p}_{m,i} = p_{m,i}\left \vert
        \Tilde{g}_{m,i} + \Tilde{\boldsymbol{G}}^{\mathsf{H}} \boldsymbol{\Theta}_m \Tilde{\boldsymbol{h}}_{m,i}
    \right \vert^2,
    \label{dm21}
\end{equation}
\begin{equation}
    \Bar{p}_{m,j} = p_{m,j}\left \vert
        \Tilde{g}_{m,j} + \Tilde{\boldsymbol{G}}^{\mathsf{H}} \boldsymbol{\Theta}_m \Tilde{\boldsymbol{h}}_{m,j}
    \right \vert^2.
    \label{dm22}
\end{equation}
\begin{proof}
    \eqref{dm} can be written as 
    \begin{equation}
        D_m = \sum_{i=1}^2 d_{m,i}^{\mathrm{}} + \sum_{i=1}^2 d_{m,i}^{\mathrm{tr}} \label{pro1proof1},
    \end{equation}
    By utilizing the intrinsic properties of logarithmic functions to the right-hand side of \eqref{pro1proof1}. Then, we can get \eqref{dm2}.
\end{proof}
\end{proposition}

\subsection{Optimizing $\boldsymbol{F}$ Given $\boldsymbol{t}$, $\boldsymbol{p}$, and $\boldsymbol{\Theta}$}
When we fix the time allocation $\dot{\boldsymbol{t}} = \{ \dot{t}_m, m \in \mathcal{T}\}$, the transmit power of each RSU $\dot{\boldsymbol{p}}= \{ \dot{p}_{m,i}, m \in \mathcal{M}, i \in \{1,2\} \}$, and the RIS phase shift coefficients $\dot{\boldsymbol{\Theta}} = \{ \dot{\boldsymbol{\Theta}}_0\ \text{and}\ \dot{\boldsymbol{\Theta}}_m, m \in \mathcal{M} \}$, problem \textbf{P2} can be simplified as
\begin{subequations} { 
\begin{flalign}
 (\textbf{P3}):\ \max_{\boldsymbol{F}, \ell} \quad & \ell  \nonumber\\
 {\rm{s.t.}}  \quad 
  &\kappa F_{m,i}^3 (T-\dot{t}_0) + \dot{t}_m \dot{p}_{m,i} \leq e_{m,i}^{\mathrm{NLEH}}, \nonumber\\
 &\quad\quad\forall m \in \mathcal{M},\ \forall i \in \{1,2 \},\label{p3a}\\
 &0 \leq F_{m,i} \leq F_{m, i, \max},\nonumber\\
 &\quad\quad\forall m\in \mathcal{M},\ \forall i \in \{1,2 \}, \label{p3b}\\
 &\ell \leq D_m,\ \forall m \in \mathcal{M}.\label{p3c}
\end{flalign}}\end{subequations}Then, the solution to problem \textbf{P3} can be given by the following proposition.
\begin{proposition}
    The solution of the CRCs allocation to problem \textbf{P3} can be expressed as
    \begin{equation}
        F_{m,i}^{\ast} = \min
        \left(
            \sqrt[\leftroot{-2}\uproot{15}3]{\frac{e_{m,i}^{\mathrm{NLEH}}-\dot{t}_m\dot{p}_{m,i}}{\kappa(T-\dot{t}_0)}},
            F_{m, i, \max}
        \right).\label{pro2}
    \end{equation}
    \begin{proof}
        It is obviously that problem \textbf{P3} is a convex optimization problem. The left-hand side of the constraint \eqref{p3a} is monotonically increasing with respect to $F_{m,i}$, and the constraints \eqref{p3b} and \eqref{p3c} are linear. Moreover, since the objective is to maximize $\ell$, and $\ell$ increases monotonically with $D_m$ (which itself is monotonically increasing with $F_{m,i}$), increasing $F_{m,i}$ will lead to a higher $\ell$. Therefore, the optimal $F_{m,i}^{\ast}$ must be attained at the boundary of the feasible set. In addition, this boundary is jointly determined by the energy constraint \eqref{p3a}. Consequently, the optimal solution is directly given by \eqref{pro2}.

    \end{proof}
\end{proposition}

\subsection{Optimizing $\boldsymbol{t}$ Given $\boldsymbol{F}$, $\boldsymbol{p}$, and $\boldsymbol{\Theta}$}
When we fix the CRCs $\dot{\boldsymbol{F}} = \{ \dot{F}_{m,i}, m \in \mathcal{T}\}, i \in \{1,2\}\}$, the transmit power of each RSU $\dot{\boldsymbol{p}}= \{ \dot{p}_{m,i}, m \in \mathcal{M}, i \in \{1,2\} \}$, and the RIS phase shift coefficients $\dot{\boldsymbol{\Theta}} = \{ \dot{\boldsymbol{\Theta}}_0\ \text{and}\ \dot{\boldsymbol{\Theta}}_m, m \in \mathcal{M} \}$, problem \textbf{P2} can be simplified as
\begin{subequations} { 
\begin{flalign}
 (\textbf{P4}):\ \max_{\boldsymbol{t}, \ell} \quad & \ell  \nonumber\\
 {\rm{s.t.}}  \quad 
  &\sum_{m=0}^M t_m \leq T,\label{p4a}\\
 &0 \leq t_m \leq T,\ \forall m \in \mathcal{T},\label{p4b}\\
 &\kappa \dot{F}_{m,i}^3 (T-t_0) + t_m \dot{p}_{m,i} \leq e_{m,i}^{\mathrm{NLEH}}, \nonumber\\
 &\quad\quad\forall m \in \mathcal{M},\ \forall i \in \{1,2 \},\label{p4c}\\
 &\ell \leq D_m,\ \forall m \in \mathcal{M}.\label{p4d}
\end{flalign}}\end{subequations}Problem \textbf{P4} can be proved as a convex optimization problem, since \textbf{P4} exhibits linearity in both the objective function and constraints. Then, we can utilize the \textsc{Yalmip}, a MATLAB Toolbox \cite{10606316}, to obtain the solution to \textbf{P4}.

\subsection{Optimizing $\boldsymbol{p}$ Given $\boldsymbol{F}$, $\boldsymbol{t}$, and $\boldsymbol{\Theta}$}
When we fix the CRCs $\dot{\boldsymbol{F}} = \{ \dot{F}_{m,i}, m \in \mathcal{T}\}, i \in \{1,2\}\}$, time allocation $\dot{\boldsymbol{t}} = \{ \dot{t}_m, m \in \mathcal{T}\}$, and the RIS phase shift coefficients $\dot{\boldsymbol{\Theta}} = \{ \dot{\boldsymbol{\Theta}}_0\ \text{and}\ \dot{\boldsymbol{\Theta}}_m, m \in \mathcal{M} \}$, problem \textbf{P2} can be simplified as
\begin{subequations} { 
\begin{flalign}
 (\textbf{P5}):\ \max_{\boldsymbol{p}, \ell} \quad & \ell  \nonumber\\
 {\rm{s.t.}}  \quad 
  &0 \leq p_{m,i} \leq P_{m, i, \max},\ \forall m\in \mathcal{M},\ \forall i \in \{1,2 \},\label{p5a}\\
 &\kappa \dot{F}_{m,i}^3 (T-\dot{t}_0) + \dot{t}_m p_{m,i} \leq e_{m,i}^{\mathrm{NLEH}}, \nonumber\\
 &\quad\quad\forall m \in \mathcal{M},\ \forall i \in \{1,2 \},\label{p5b}\\
 &\ell \leq D_m,\ \forall m \in \mathcal{M}.\label{p5c}
\end{flalign}}\end{subequations}
Observing that the right-hand side of constraint \eqref{p5c} is concave in the transmit power of each RSU, and the objective function and all remaining constraints are convex, problem \textbf{P5} is a convex optimization problem.
Hence, \textbf{P5} can be solved utilizing conventional convex optimization tools, such as \textsc{CVX}.
Besides, to obtain more insights, we also analyze the closed-form expressions of the optimal solution to the power allocation, which is shown in Theorem \ref{Theorem1}.

\begin{theorem}\label{Theorem1}
The closed-form expression of the optimal solution to \textbf{P5} can be expressed as
\begin{equation}
\begin{split}
{p}_{m,i}^* = \frac{\ln(2)\beta_{m,i}B}{\lambda_{m,i}}-\frac{\sigma_m^2 + \dot{\Bar{p}}_{m,j}}{\left \vert
        \Tilde{g}_{m,i} + \Tilde{\boldsymbol{G}}^{\mathsf{H}} \dot{\boldsymbol{\Theta}}_m \Tilde{\boldsymbol{h}}_{m,i}
    \right \vert^2},
\end{split}
\end{equation} where $\lambda_{m,i}$ and $\beta_{m,i}$ are Lagrange multipliers associated with constraints \eqref{p5b} and \eqref{p5c}, respectively, and $\dot{\Bar{p}}_{m,j}$ is given by
\begin{equation}
    \dot{\Bar{p}}_{m,j} = {p}_{m,j}\left \vert
        \Tilde{g}_{m,j} + \Tilde{\boldsymbol{G}}^{\mathsf{H}} \dot{\boldsymbol{\Theta}}_m \Tilde{\boldsymbol{h}}_{m,j}
    \right \vert^2.
\end{equation}

\begin{proof}
Please refer to Appendix \ref{appendice1}.
\end{proof}
\end{theorem} 

\subsection{Optimizing $\boldsymbol{\Theta}$ Given $\boldsymbol{t}$, $\boldsymbol{p}$, and $\boldsymbol{F}$}
When we fix the time allocation $\dot{\boldsymbol{t}} = \{ \dot{t}_m, m \in \mathcal{T}\}$, the transmit power of each RSU $\dot{\boldsymbol{p}}= \{ \dot{p}_{m,i}, m \in \mathcal{M}, i \in \{1,2\} \}$, and the CRCs $\dot{\boldsymbol{F}} = \{ \dot{F}_{m,i}, m \in \mathcal{M}, i \in \{1,2\} \}$, problem \textbf{P2} can be simplified as
\begin{subequations} { 
\begin{flalign}
 (\textbf{P6}):\ \max_{\boldsymbol{\Theta}, \ell} \quad & \ell  \nonumber\\
 {\rm{s.t.}}  \quad 
 &e_{m,i}^{\mathrm{proc}} + e_{m,i}^{\mathrm{tr}} \leq e_{m,i}^{\mathrm{NLEH}}, \nonumber\\
 &\quad\quad\forall m \in \mathcal{M},\ \forall i \in \{1,2 \},\label{p6a}\\
 & 0\leq \theta_{0,n} \le 2\pi,\ \forall n \in \mathcal{N},\label{p6b}\\
 &0\leq \theta_{m,n} \le 2\pi,\ \forall n \in \mathcal{N},\ \forall m \in \mathcal{M},\label{p6c}\\
 &\ell \leq D_m,\ \forall m \in \mathcal{M}.\label{p6d}
\end{flalign}}\end{subequations}
Let $\boldsymbol{H}_{m,i} = [\mathsf{diag} \{ \boldsymbol{h}_{m,i}\}\boldsymbol{G}; g_{m,i}]$, and $\boldsymbol{v}_0 = [\mathrm{e}^{\mathrm{j}\theta_{0,1}},\ldots,\mathrm{e}^{\mathrm{j}\theta_{0,N}}, 1]^{\mathsf{H}}$, the equivalent channel gain from the DC to the RSU $m_i$ can be rewritten as
\begin{equation}
    \boldsymbol{h}_{m,i}^{\mathsf{H}} \boldsymbol{\Theta}_0 \boldsymbol{G} + g_{m,i} = \boldsymbol{v}_0^{\mathsf{H}} \boldsymbol{H}_{m,i}.
\end{equation}
Similarly, the equivalent channel gain for the DC data harvesting from the RSU $m_i$ can be reformulated as
\begin{equation}
    \Tilde{g}_{m,i} + \Tilde{\boldsymbol{G}}^{\mathsf{H}} \boldsymbol{\Theta}_m \Tilde{\boldsymbol{h}}_{m,i} = \boldsymbol{v}_{m}^{\mathsf{H}} \Tilde{\boldsymbol{H}}_{m,i},
\end{equation}
where $\Tilde{\boldsymbol{H}}_{m,i} = [\mathsf{diag} \{ \Tilde{\boldsymbol{h}}_{m,i}\}\Tilde{\boldsymbol{G}}; \Tilde{g}_{m,i}]$, and $\boldsymbol{v}_{m} = [\mathrm{e}^{\mathrm{j}\theta_{m,1}},\ldots,\mathrm{e}^{\mathrm{j}\theta_{m,N}}, 1]^{\mathsf{H}}$.

Furthermore, through inequality transformation, constraint \eqref{p6a} can be reformulated as
\begin{equation}
    \frac{\Psi_{m,i}}{a_{m,i}} + b_{m,i} \leq
    \vert \boldsymbol{v}_0^{\mathsf{H}} \boldsymbol{H}_{m,i} \vert^2,\quad\forall m \in \mathcal{M},\ \forall i \in \{1,2 \},
\end{equation}
where $\Psi_{m,i} = \ln\frac{(1-\Omega_{m,i})(e_{m,i}^{\mathrm{proc}} + e_{m,i}^{\mathrm{tr}}) + \dot{t}_0\Lambda_{m,i}\Omega_{m,i}}{(1-\Omega_{m,i})(\dot{t}_0\Lambda_{m,i}-e_{m,i}^{\mathrm{proc}} - e_{m,i}^{\mathrm{tr}}))}$.

We denote $\Bar{\boldsymbol{H}}_{m,i} = \boldsymbol{H}_{m,i}\boldsymbol{H}_{m,i}^{\mathsf{H}}$, and $\hat{\boldsymbol{H}}_{m,i} = \Tilde{\boldsymbol{H}}_{m,i} \Tilde{\boldsymbol{H}}_{m,i}^{\mathsf{H}}$.
Then, we denote $\boldsymbol{V}_{t} = \boldsymbol{v}_t \boldsymbol{v}_t^{\mathsf{H}}, \forall t \in \mathcal{T}$. Obviously, we have $\mathsf{rank}(\boldsymbol{V}_{t}) = 1$, and $\boldsymbol{V}_{t} \succcurlyeq \boldsymbol{0}$.
Based on the above transformation, problem \textbf{P6} can be reformulated as
\begin{subequations} { 
\begin{flalign}
 (\textbf{P6.1}):\ \max_{\boldsymbol{V}, \ell} \quad & \ell  \nonumber\\
 {\rm{s.t.}}  \quad 
 &\frac{\Psi_{m,i}}{a_{m,i}} + b_{m,i} \leq
    \vert \boldsymbol{v}_0^{\mathsf{H}} \boldsymbol{H}_{m,i} \vert^2, \nonumber\\
 &\quad\quad\forall m \in \mathcal{M},\ \forall i \in \{1,2 \},\label{p7a}\\
 & \mathsf{rank}(\boldsymbol{V}_{t}) = 1,\ \forall t \in \mathcal{T},\label{p7b}\\
 & \boldsymbol{V}_{t} \succcurlyeq \boldsymbol{0},\ \forall t \in \mathcal{T},\label{p7c}\\
 &[\boldsymbol{V}_{t}]_{n,n}=1,\ \forall t \in \mathcal{T},\ n = 1,\ldots,N,N+1,\label{p7d}\\
 &\Tilde{\ell} \leq \dot{p}_{m,1} \mathrm{tr}(\hat{\boldsymbol{H}}_{m,1}\boldsymbol{V}_{m}) + \dot{p}_{m,2}\mathrm{tr}(\hat{\boldsymbol{H}}_{m,2}\boldsymbol{V}_{m}), \nonumber\\
 &\quad\quad\ \forall m \in \mathcal{M},\label{p7e}
\end{flalign}}\end{subequations}
where $\boldsymbol{V} = \{ \boldsymbol{V}_{t}, \forall t \in \mathcal{T} \}$, and $\Tilde{\ell}$ is given by
\begin{equation}
    \Tilde{\ell} = \left (
        2^{\frac{D_m - \sum_{i=1}^2d_{m,i}}{B\dot{t}_m}}-1
    \right ) \sigma_m^2.
\end{equation}
\begin{remark}
    It is noted that the sole source of non-convexity in problem \textbf{P6.1} stems from the rank-one constraint \eqref{p7b}. While prior works predominantly employ semi-definite relaxation (SDR) to address this constraint \cite{10288203,9930110,10721288}, such approaches inherently suffer from a fundamental limitation: the relaxed solutions often fail to satisfy the original rank-one requirement, leading to suboptimal or even infeasible outcomes. To mitigate this issue, we adopt the SROCR method \cite{9599533, 10807234}, which progressively relaxes the rank-one constraint through iterative refinement. Unlike the SDR method, the SROCR implements a controlled relaxation mechanism that preserves traceable rank-one characteristics while navigating toward locally optimal solutions, thereby balancing feasibility and solution quality.
\end{remark}

Specifically, during the $q$-th iteration, we formulate a relaxed problem \textbf{P7}.$q$ by utilizing the solution $\boldsymbol{V}^{(q-1)}$ from the $(q-1)$-th iteration as an approximation for the subsequent iteration. Here, the relaxed problem \textbf{P7}.$q$ in the $q$-th iteration is formulated as
\begin{subequations} { 
\begin{flalign}
 (\textbf{P7.}q):\ \max_{\boldsymbol{V}, \ell} \quad & \ell  \nonumber\\
 {\rm{s.t.}}  \quad 
 &\eqref{p7a},\ \eqref{p7c}\text{-}\eqref{p7e},\nonumber\\
 &\left(\boldsymbol{u}_t^{(q-1)}\right)^{\mathsf{H}}\boldsymbol{V}_t\boldsymbol{u}_t^{(q-1)} \geq \chi^{(q-1)} \mathrm{tr} (\boldsymbol{V}_t), \nonumber\\
 &\quad\quad\ \forall t \in \mathcal{T},\label{p8a}
\end{flalign}}\end{subequations}
where $\boldsymbol{u}_t^{(q-1)}$ denotes the principal eigenvector of $\boldsymbol{V}_t$, and $ \chi^{(q-1)} $ is an introduced relaxation parameter. Problem \textbf{P7.}$q$ is already convex. The relaxation from \eqref{p7b} to \eqref{p8a} becomes tighter and tighter as $ \chi^{(q)} $ increases from 0 to 1 over iterations. Specifically, when $ \chi^{(q)} = 1 $, \eqref{p8a} is equivalent to \eqref{p7b}, and thus the solution obtained by solving \textbf{P7.}$q$ also serves as a solution to \textbf{P6.1}.

\subsection{Algorithm Analysis}
The overall AO algorithm for solving \textbf{P1} is summarized in Algorithm \ref{algoverall},
where the three sub-problems are solved in lines 2–5 by the AO manner until either the maximum number of iterations or the precision threshold is met.

\begin{algorithm} \scriptsize
	\caption{AO Algorithm for Solving \textbf{P1}}
	\label{algoverall}
	\begin{algorithmic}[1]
		\REQUIRE {An initial feasible solution ${\boldsymbol{F}}^{i}$, ${\boldsymbol{t}}^{i}$, ${\boldsymbol{p}}^{i}$, and ${\boldsymbol{\Theta}}^{i}$;}\\
		\textbf{Initialize:} the iteration number $i=0$;\\
		\textbf{Initialize:} the iteration precision threshold $\epsilon$;\\
            \textbf{Initialize:} the maximum number of iterations $i_{\max}$;\\
		\REPEAT
		\STATE Solving problem \textbf{P3} and get the solution ${\boldsymbol{F}}^{i+1}$, when given ${\boldsymbol{t}}^{i}$, ${\boldsymbol{p}}^{i}$, and ${\boldsymbol{\Theta}}^{i}$;\\
		\STATE Solving problem \textbf{P4} and get the solution ${\boldsymbol{t}}^{i+1}$, when given ${\boldsymbol{F}}^{i+1}$, ${\boldsymbol{p}}^{i}$, and ${\boldsymbol{\Theta}}^{i}$;\\
            \STATE Solving problem \textbf{P5} and get the solution ${\boldsymbol{p}}^{i+1}$, when given ${\boldsymbol{F}}^{i+1}$, ${\boldsymbol{t}}^{i+1}$, and ${\boldsymbol{\Theta}}^{i}$;\\
		\STATE Solving problem \textbf{P6} and get the solution ${\boldsymbol{\Theta}}^{i+1}$, when given ${\boldsymbol{F}}^{i+1}$, ${\boldsymbol{t}}^{i+1}$, and ${\boldsymbol{p}}^{i+1}$;\\
            \STATE Calculating the objective value of problem \textbf{P1} as $\Tilde{D}^{i+1}$;
		\STATE Set $i=i+1$; \\
		\UNTIL{$|\Tilde{D}^{i}-\Tilde{D}^{i-1}|$$<$$\epsilon$} or $i = i_{\max}$;\\
		\ENSURE {The optimized solution ${\boldsymbol{F}}^{i}$, ${\boldsymbol{t}}^{i}$, ${\boldsymbol{p}}^{i}$, and $\bold{{\boldsymbol{\Theta}}}^{i}$.}\\
	\end{algorithmic}
\end{algorithm}

\paragraph{Convergence Analysis}
The convergence of the proposed Algorithm \ref{algoverall} is analyzed in the subsequent proposition.

\begin{proposition}
\label{theorem 3}
    The proposed scheme ensures that the objective function value of problem \textbf{P2} is non-decreasing throughout the iterations and finally converges.
    
\begin{proof}
    Please refer to Appendix \ref{appendice2}.
\end{proof}
\end{proposition}

\paragraph{Computational Complexity Analysis}
The computational complexity of the proposed scheme primarily depends on the solution procedures of sub-problems \textbf{P4} and \textbf{P6.1}. For sub-problem \textbf{P4}, the number of variables is \(a_1 = (M + 2)\), and the number of constraints is \(b_1 = 5M + 3\). Under the condition of a given solution accuracy \(\varepsilon_1 > 0\), its worst-case computational complexity is \(I_1 = {\cal O}((a_1^2b_1 + a_1^3)b_1^{1/2}\log(1/\varepsilon_1))\) \cite{Dai2018}.
Similarly, the computational complexity of the sub-problem \textbf{P6.1} can be expressed as \(I_2 = {\cal O}((a_2^2b_2 + a_2^3)b_2^{1/2}\log(1/\varepsilon_2))\), where the number of variables \(a_2=(N + 1)^2(M + 1)\) and the number of constraints \(b_2 = 4M + 1+(M + 1)(N + 1)\). Let the number of iterations of the overall solving process and the SROCR method be \(I_{AO}\) and \(I_{\text{SROCR}}\), respectively. The overall computational complexity of the proposed scheme can be expressed as \({\cal O}(I_{AO}(I_1 + I_{\text{SROCR}}I_2))\).

\begin{table}[!htbp] \scriptsize
\centering
	\caption{Main parameter settings.}
	\label{table:1}
\begin{tabular}{cc}
\toprule
\centering
Parameters& Value\\
                \midrule
			\centering System bandwidth ($ B $) & $ 100 $ $\mathrm{KHz} $ \\
			\centering Maximum transmit power of DC ($ P_{\max} $) & $ 34 $ $ \mathrm{dBm} $ \\
			\centering Number of reflecting elements ($ N $) & $ 30 $ \\
			\centering The CRC capability of the RSUs ($ F_{m,i,\max} $) & $ 1.8 \times 10^7 $ $ \mathrm{cycles/s} $ \\
			\centering Time block length ($ T $) & $ 1.0 $ s \\
			\centering The complexity of tasks ($ f_{m,i} $) & $ 1000 $ $ \mathrm{cycles/bit} $ \\
			\centering The circuit capacitance coefficient ($ \kappa $) & $ 10^{-28} $ \\
			\centering Noise power ($ \sigma_m^2 $) & $ -8 $ $ \mathrm{dBm} $ \\
			\centering Path-loss for $ {\boldsymbol{G}} $, $ \Tilde{\boldsymbol{G}} $, $ {\boldsymbol{h}}_{m,i} $, and $ \Tilde{\boldsymbol{h}}_{m,i} $ & $ 35.6+22\ \mathrm{lg}\left( d \right) $ \\
			\centering Path-loss for $ {g}_{m,i} $ and $ \Tilde{g}_{m,i,} $ & $ 32.6+36.7\ \mathrm{lg}\left( d \right) $ \\
			\bottomrule
	\end{tabular}
	\vspace{-1em}
\end{table}

\section{Numerical Results}
This section presents simulation results aimed at assessing the performance of the proposed scheme in the ICTS. The simulation setup includes a single-antenna DC situated at the coordinates \((0,0)\), and an RIS placed at \((75,30)\). Four RSUs are organized into two pairs, each pair being distributed within circular regions with a radius of 10 meters. The centers of these circular regions are positioned at \((65, 10)\) and \((85, 10)\) respectively. The key simulation parameters are detailed in Table \ref{table:1}.
To enable a thorough performance comparison, three baseline schemes are also simulated.

\begin{itemize}
        \item Only TDMA Scheme: In the uplink, RSUs transmit processed data to the DC using only the TDMA technique.
	\item Random Phase Scheme: The RIS phase shifts are randomly assigned.
	\item Without RIS Scheme: Removing the RIS from the system, i.e., the system communicates only through the direct link.

\end{itemize}
\begin{figure}[t]
	\centering
        \includegraphics[width=0.782\linewidth]{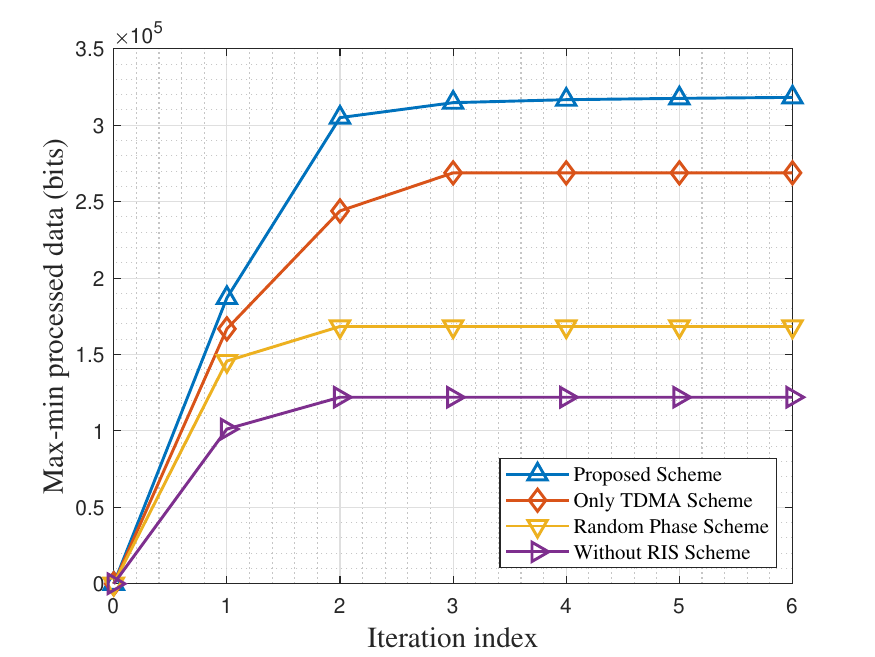}
	\caption{The convergence behavior of all the schemes.}
	\label{convergence}
\end{figure}

Fig. \ref{convergence} presents a comparison of the convergence behavior across all schemes, and it clearly demonstrates that the proposed scheme outperforms all baseline schemes in terms of max-min processed data. In contrast to the random phase scheme, the phase shift of the RIS under the proposed scheme is optimized, which in turn results in a higher volume of processed data. When compared with the only TDMA scheme, RSMA stands out by concurrently encoding multiple data streams through rate-splitting, thereby facilitating improved interference management and multi-user cooperation. Finally, the other three schemes achieve a higher level of max-min processed data than the without RIS scheme, as they possess an additional link that enhances communication between the DC and RSUs, ultimately boosting the overall performance.
\begin{figure}[t]
	\centering
        \includegraphics[width=0.782\linewidth]{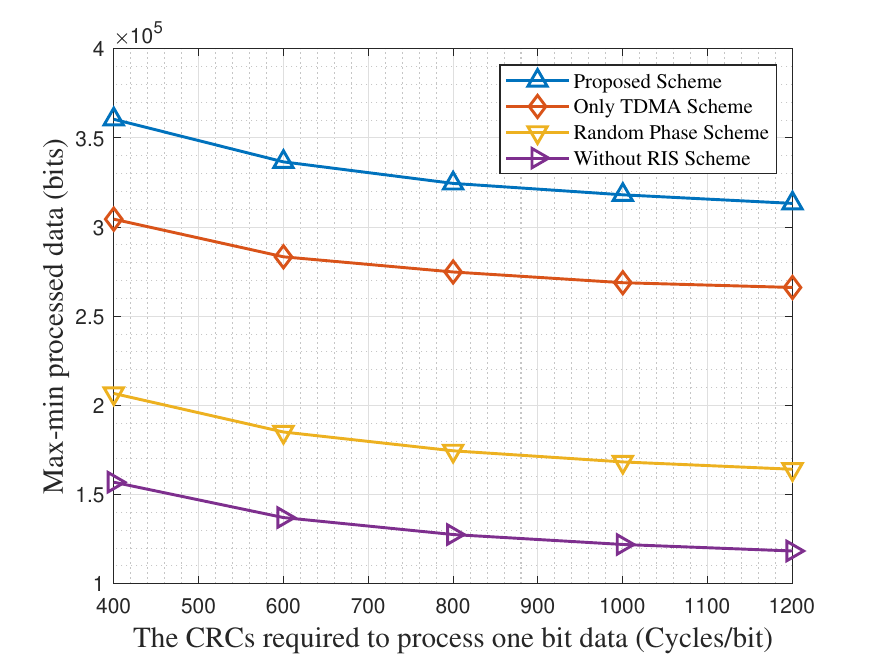}
	\caption{Max-min processed data versus the CRCs required for processing 1-bit task.}
	\label{cycle}
\end{figure}

Fig. \ref{cycle} explores the effect of the CRCs needed to process 1-bit task data (i.e., $ f_{m,i} $) on the max-min processed data of RSUs. Across all schemes, the max-min processed data exhibits a decreasing trend as $ f_{m,i} $ increases. The reason behind this lies in the fact that, according to (5), a higher $ f_{m,i} $ necessitates more CRCs for processing 1-bit of task data. This higher demand leads to a reduction in the number of task bits that RSUs can process, and as a result, the overall max-min processed data declines.
\begin{figure}[t]
	\centering
        \includegraphics[width=0.782\linewidth]{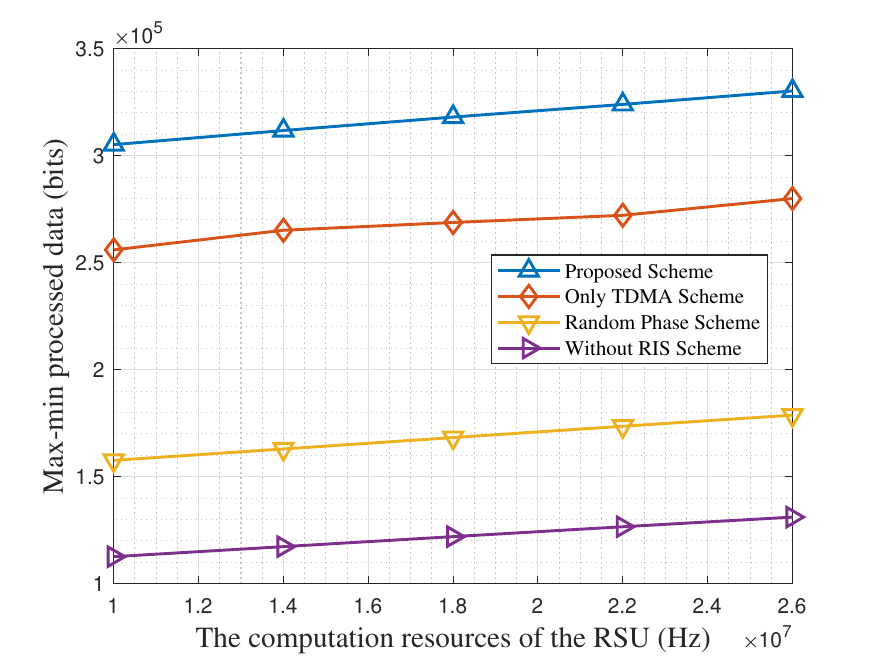}
	\caption{Max-min processed data versus the RSUs' computation resources.}
	\label{frequency}
\end{figure}

Fig. \ref{frequency} illustrates the relationship between the max-min processed data and the RSUs' computation resources (i.e., $ F_{m,i,\text{max}} $) across different schemes. A striking observation is that for all schemes, an increase in $ F_{m,i,\text{max}} $ leads to a significant rise in the max-min processed data. This trend is directly explained by the computational model: higher computation resources enables RSUs to execute more task bits within the same time frame, as described by (5). 
\begin{figure}[t]
	\centering
        \includegraphics[width=0.782\linewidth]{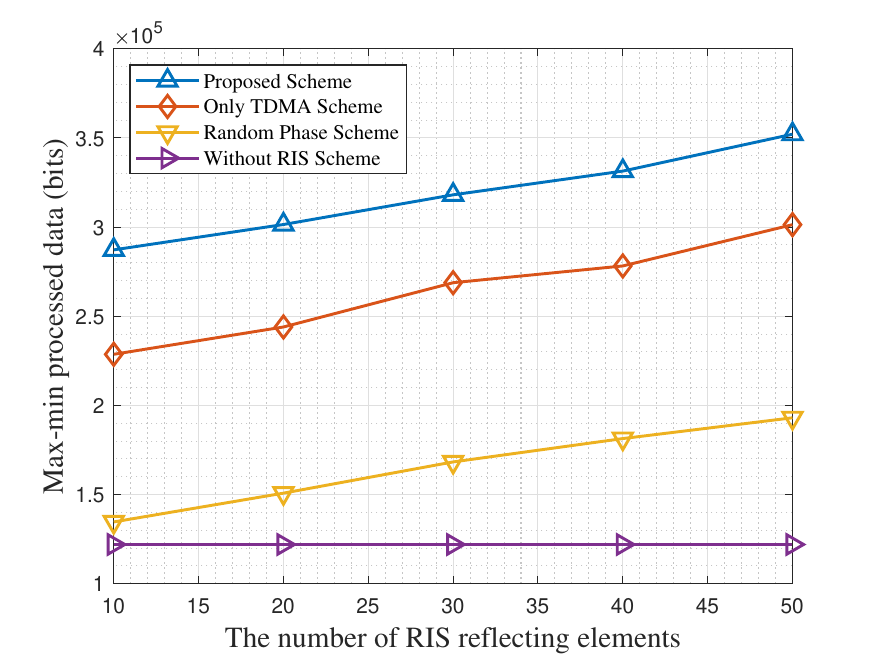}
	\caption{Max-min processed data versus the number of RIS reflecting elements.}
	\label{rise}
\end{figure}

Fig. \ref{rise} illustrates how the number of RIS reflecting elements affects the max-min processed data. It is evident that the max-min processed data increases in tandem with a growing number of reflecting elements. This phenomenon can be attributed to the fact that a larger number of RIS elements directly enhances the system's channel gain. Specifically, more reflecting elements aggregate stronger signal reflections, which collectively boost the signal strength between the DC and RSUs. As a result, the enhanced channel conditions lead to a higher volume of max-min processed data.
\begin{figure}[t]
	\centering
        \includegraphics[width=0.782\linewidth]{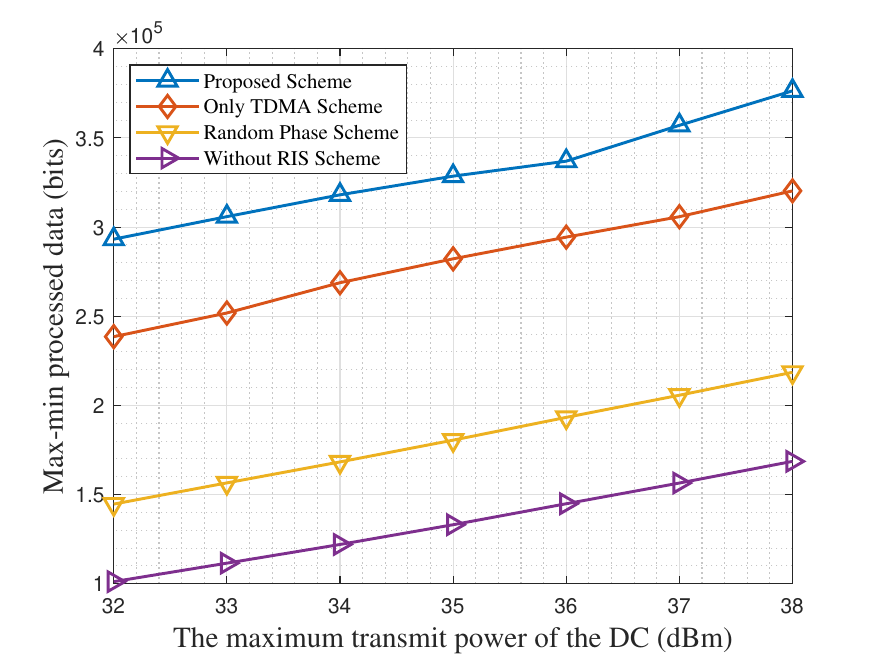}
	\caption{Max-min processed data versus the DC's transmit power.}
	\label{trasmitp}
\end{figure}

Fig. \ref{trasmitp} shows the variation of max-min processed data with the DC's maximum transmit power. Eq. \eqref{recepow} confirms that the RF power received by the RSUs is proportional to the DC's transmit power, which means that a higher transmit power directly leads to an increase in the energy harvested by the RSUs. Therefore, the RSUs can utilize the extra harvested energy to process more data, and as a result, the max-min processed data increases steadily as the maximum transmit power of the DC rises.

\section{Conclusion}
This paper presents a RIS-enabled framework for the ICTS, focusing on enhancing the data collection capabilities. To further improve the system performance, a novel hybrid transmission protocol merging RSMA with TDMA is proposed. The objective is to maximize the minimal processed data for each RSU pair through the joint optimization of the RIS downlink/uplink phase shift coefficients, the time slot allocation, the transmit power of DC and RSUs, and the RSU CRCs. An efficient algorithm leveraging AO and SROCR is proposed to solve the formulated problem. The simulation results confirm that the proposed scheme performs better than baseline schemes.

\begin{appendices}
\section{Proof of Theorem \ref{Theorem1}}\label{appendice1}
\begin{proof} The Lagrangian function of problem \textbf{P5} can be expressed as
\begin{equation} {
\begin{split}
\mathcal{L}(&\boldsymbol{p},\ell,\boldsymbol{\lambda},\boldsymbol{\beta} )\\
& = -\ell + \sum_{m=1}^M \sum_{i=1}^2 \lambda_{m,i} \left( 
    \kappa \dot{F}_{m,i}^3 (T-\dot{t}_0) + \dot{t}_m \dot{p}_{m,i} - e_{m,i}^{\mathrm{NLEH}}
\right)\\
& \quad\ + \sum_{m=1}^M \sum_{i=1}^2 \beta_{m,i} (\ell - D_m),
\end{split} }
\end{equation}
where $\boldsymbol{\lambda} = \left\{ \lambda_{m,i}, \forall m \in \mathcal{M}, i \in \{1,2\} \right\}$ and $\boldsymbol{\beta} = \left\{ \beta_{m,i}, \forall m \in \mathcal{M}, i \in \{1,2\} \right\}$ are the dual variables associated with the corresponding constraints \eqref{p5b} and \eqref{p5c}, respectively.
Then, the Karush-Kuhn-Tucker (KKT) conditions can be expressed as
\begin{equation}
\label{kkt1} {
\begin{split}
\frac{\partial \mathcal{L}}{\partial p_{m,i}} &= \lambda_{m,i}\dot{t}_m \\ & - \beta_{m,i}\frac{\ln(2)B\dot{t}_m\left \vert
        \Tilde{g}_{m,i} + \Tilde{\boldsymbol{G}}^{\mathsf{H}} \dot{\boldsymbol{\Theta}}_m \Tilde{\boldsymbol{h}}_{m,i}
    \right \vert^2}{\dot{\Bar{p}}_{m,i}+\dot{\Bar{p}}_{m,j}+\sigma_m^2}
=0,\\
&\quad\quad\quad\quad\quad\quad\quad\quad
\forall m \in \mathcal{M},\ i,j \in 
\{1,2\},\ \text{and}\ i \neq j,
\end{split} }
\end{equation}
\begin{equation} {
\begin{split}
\frac{\partial \mathcal{L}}{\partial \ell} =
-1 + \sum_{m=1}^M \sum_{i=1}^2 \beta_{m,i} = 0,
\end{split}}
\end{equation}
\begin{equation} {
\begin{split}
&\lambda_{m,i} \left( 
   \kappa \dot{F}_{m,i}^3 (T-\dot{t}_0) + \dot{t}_m p_{m,i} - e_{m,i}^{\mathrm{NLEH}}
\right) = 0,\\&
\quad\quad\quad\quad
\forall m \in \mathcal{M},\ i,j \in 
\{1,2\},\ \text{and}\ i \neq j,
\end{split} }
\end{equation}
\begin{equation}
\label{kkt4} {
\begin{split}
&\beta_{m,i} (\ell - D_m) = 0,\\
&\quad\quad
\forall m \in \mathcal{M},\ i,j \in 
\{1,2\},\ \text{and}\ i \neq j.
\end{split} }
\end{equation}
 Then, the optimal power allocation can be obtained by combining \eqref{kkt1}, \eqref{dm21}, and \eqref{dm22}.
\end{proof}

\section{Proof of Proposition \ref{theorem 3}}\label{appendice2}
\begin{proof} Firstly, define $ {R_{{P_2}}}({\dot{\boldsymbol{F}}}^{(z)}, {\dot{\boldsymbol{t}}}^{(z)}, {\dot{\boldsymbol{p}}}^{(z)}, {\dot{\boldsymbol{\Theta}}}^{(z)}) $, $ {R_{{P_3}}}({\dot{\boldsymbol{F}}}^{(z)}, \\ {\dot{\boldsymbol{t}}}^{(z)}, {\dot{\boldsymbol{p}}}^{(z)}, {\dot{\boldsymbol{\Theta}}}^{(z)}) $, $ {R_{{P_4}}}({\dot{\boldsymbol{F}}}^{(z)}, {\dot{\boldsymbol{t}}}^{(z)}, {\dot{\boldsymbol{p}}}^{(z)}, {\dot{\boldsymbol{\Theta}}}^{(z)}) $, $ {R_{{P_5}}}({\dot{\boldsymbol{F}}}^{(z)}, {\dot{\boldsymbol{t}}}^{(z)}, \\{\dot{\boldsymbol{p}}}^{(z)}, {\dot{\boldsymbol{\Theta}}}^{(z)}) $ and $ {R_{{P_7}}}({\dot{\boldsymbol{F}}}^{(z)}, {\dot{\boldsymbol{t}}}^{(z)}, {\dot{\boldsymbol{p}}}^{(z)}, {\dot{\boldsymbol{\Theta}}}^{(z)}) $ as the objective function values for problems \textbf{P2}, \textbf{P3}, \textbf{P4}, \textbf{P5}, and \textbf{P7}.$q$ in the $ z $-th iteration, respectively. Next, for the CRCs optimization problem \textbf{P3} in the $ (z+1) $-th iteration, the following holds
	\begin{align}
		&{R_{{P_2}}}({\dot{\boldsymbol{F}}}^{(z)}, {\dot{\boldsymbol{t}}}^{(z)}, {\dot{\boldsymbol{p}}}^{(z)}, {\dot{\boldsymbol{\Theta}}}^{(z)}) \notag \\
		&\mathop = \limits^{(a)} {R_{{P_3}}}({\dot{\boldsymbol{F}}}^{(z)}, {\dot{\boldsymbol{t}}}^{(z)}, {\dot{\boldsymbol{p}}}^{(z)}, {\dot{\boldsymbol{\Theta}}}^{(z)}) \notag \\
		&\mathop \le \limits^{(b)} {R_{{P_3}}}({\dot{\boldsymbol{F}}}^{(z+1)}, {\dot{\boldsymbol{t}}}^{(z)}, {\dot{\boldsymbol{p}}}^{(z)}, {\dot{\boldsymbol{\Theta}}}^{(z)}) \notag \\
		&= {R_{{P_2}}}({\dot{\boldsymbol{F}}}^{(z+1)}, {\dot{\boldsymbol{t}}}^{(z)}, {\dot{\boldsymbol{p}}}^{(z)}, {\dot{\boldsymbol{\Theta}}}^{(z)}), \label{37}
	\end{align}
	where $ (a) $ is valid because \textbf{P2} and \textbf{P3} have identical optimal solution, and $ (b) $ arises since $ {\dot{\boldsymbol{F}}}^{(z+1)} $ denotes the optimal solution to \textbf{P3} in the $ (z+1) $-th iteration. 
	
 Applying similar reasoning to the time allocation optimization problem \textbf{P4} in the $ (z+1) $-th iteration, we have
	\begin{align}
		&{R_{{P_2}}}({\dot{\boldsymbol{F}}}^{(z+1)}, {\dot{\boldsymbol{t}}}^{(z)}, {\dot{\boldsymbol{p}}}^{(z)}, {\dot{\boldsymbol{\Theta}}}^{(z)}) \notag \\
		&\mathop = \limits^{(a)} {R_{{P_4}}}({\dot{\boldsymbol{F}}}^{(z+1)}, {\dot{\boldsymbol{t}}}^{(z)}, {\dot{\boldsymbol{p}}}^{(z)}, {\dot{\boldsymbol{\Theta}}}^{(z)}) \notag \\
		&\mathop \le \limits^{(b)} {R_{{P_4}}}({\dot{\boldsymbol{F}}}^{(z+1)}, {\dot{\boldsymbol{t}}}^{(z+1)}, {\dot{\boldsymbol{p}}}^{(z)}, {\dot{\boldsymbol{\Theta}}}^{(z)}) \notag \\
		&= {R_{{P_2}}}({\dot{\boldsymbol{F}}}^{(z+1)}, {\dot{\boldsymbol{t}}}^{(z+1)}, {\dot{\boldsymbol{p}}}^{(z)}, {\dot{\boldsymbol{\Theta}}}^{(z)}). \label{38}
	\end{align}
    
 For the transmit power optimization problem \textbf{P5}, we obtain
	\begin{align}
		&{R_{{P_2}}}({\dot{\boldsymbol{F}}}^{(z+1)}, {\dot{\boldsymbol{t}}}^{(z+1)}, {\dot{\boldsymbol{p}}}^{(z)}, {\dot{\boldsymbol{\Theta}}}^{(z)}) \notag \\
		&\mathop = \limits^{(a)} {R_{{P_5}}}({\dot{\boldsymbol{F}}}^{(z+1)}, {\dot{\boldsymbol{t}}}^{(z+1)}, {\dot{\boldsymbol{p}}}^{(z)}, {\dot{\boldsymbol{\Theta}}}^{(z)}) \notag \\
		&\mathop \le \limits^{(b)} {R_{{P_5}}}({\dot{\boldsymbol{F}}}^{(z+1)}, {\dot{\boldsymbol{t}}}^{(z+1)}, {\dot{\boldsymbol{p}}}^{(z+1)}, {\dot{\boldsymbol{\Theta}}}^{(z)}) \notag \\
		&= {R_{{P_2}}}({\dot{\boldsymbol{F}}}^{(z+1)}, {\dot{\boldsymbol{t}}}^{(z+1)}, {\dot{\boldsymbol{p}}}^{(z+1)}, {\dot{\boldsymbol{\Theta}}}^{(z)}). \label{39}
	\end{align}
        
 For the RIS phase shift coefficients optimization problem \textbf{P7}.$q$ in the $ (z+1) $-th iteration, we have
        \begin{align}
		&{R_{{P_2}}}({\dot{\boldsymbol{F}}}^{(z+1)}, {\dot{\boldsymbol{t}}}^{(z+1)}, {\dot{\boldsymbol{p}}}^{(z+1)}, {\dot{\boldsymbol{\Theta}}}^{(z)}) \notag \\
		&\mathop = \limits^{(a)} {R_{{P_7}}}({\dot{\boldsymbol{F}}}^{(z+1)}, {\dot{\boldsymbol{t}}}^{(z+1)}, {\dot{\boldsymbol{p}}}^{(z+1)}, {\dot{\boldsymbol{\Theta}}}^{(z)}) \notag \\
		&\mathop \le \limits^{(b)} {R_{{P_7}}}({\dot{\boldsymbol{F}}}^{(z+1)}, {\dot{\boldsymbol{t}}}^{(z+1)}, {\dot{\boldsymbol{p}}}^{(z+1)}, {\dot{\boldsymbol{\Theta}}}^{(z+1)}) \notag \\
		&= {R_{{P_2}}}({\dot{\boldsymbol{F}}}^{(z+1)}, {\dot{\boldsymbol{t}}}^{(z+1)}, {\dot{\boldsymbol{p}}}^{(z+1)}, {\dot{\boldsymbol{\Theta}}}^{(z+1)}). \label{40}
	\end{align}
    
 Combining (\ref{37})-(\ref{40}), we conclude
	\begin{align}
		&{R_{{P_2}}}({\dot{\boldsymbol{F}}}^{(z)},        {\dot{\boldsymbol{t}}}^{(z)}, {\dot{\boldsymbol{p}}}^{(z)}, {\dot{\boldsymbol{\Theta}}}^{(z)}) \notag \\
            &\le {R_{{P_2}}}({\dot{\boldsymbol{F}}}^{(z+1)}, {\dot{\boldsymbol{t}}}^{(z+1)}, {\dot{\boldsymbol{p}}}^{(z+1)}, {\dot{\boldsymbol{\Theta}}}^{(z+1)}),
	\end{align}
	which indicates that the objective function of \textbf{P2} is non-decreasing during iterations. Since the variables in \textbf{P2} are bounded by problem constraints, the objective function finally converges.
\end{proof}

\end{appendices}

\bibliographystyle{IEEEtran}
\bibliography{ref}
\end{document}